\documentstyle[prl,aps,twocolumn,float,epsf]{revtex}  

\tightenlines
\begin{document}

\twocolumn[\hsize\textwidth\columnwidth\hsize\csname 
@twocolumnfalse\endcsname

\title{High Temperature Thermodynamics of the Ferromagnetic Kondo-Lattice
Model}

\author{ H. R\"oder $^1$,
R. R. P. Singh $^2$, J. Zang$^1$}
\address{${}^1$ Los Alamos National Laboratory, Los Alamos \\
${}^2$Department of Physics,                                    
University of California, Davis, California 95616}

\date{\today}

\maketitle 

\begin{abstract}
We present a high temperature series expansion for the ferromagnetic
Kondo lattice model in the large coupling limit, which is used to
model CMR perovskites. Our results show the expected cross--over
to Curie--Wei{\ss} behavior at a temperature related to the bandwidth.
Estimates for the magnetic transition temperatures
are in the experimentally observed range. The 
compressibility shows that the high temperature charge excitations 
can be modeled by spinless fermions. The CMR effect itself,
however, warrants the inclusion of dynamic effects and cannot be 
explained by a static calculation.
\end{abstract}                                                              
\phantom{.}
]
\narrowtext


The discovery of colossal magnetoresistance (CMR) in doped
rare-earth manganites has attracted considerable attention
\cite{Kus89,Cha93,vonH93}.
The double-exchange (DE) model \cite{Zen51} has long been considered appropriate
for describing these systems. In this model spins are associated
with localized $t_{2g}$ electrons, which are coupled ferromagnetically to
itinerant $e_g$ electrons through a large Hunds coupling ($J_H$).
The kinetic energy of these itinerant electrons may be lowered
by a ferromagnetic alignment of the  $t_{2g}$ electron spins resulting
in an effective ferromagnetic coupling.

Most studies of the DE model in the large $J_H$ limit
begin by mapping it
to a spinless fermion model, where the hopping matrix
elements depend on the relative orientation of neighboring 
spins \cite{And55,DeG60,Kub72}. Then various types of mean--field
theories are used to solve the resulting many--body
problem \cite{Kub72,Furu94,Millis95,Roder96}. It was pointed out by
Millis et al.\cite{Millis95} that in order to obtain the experimental value
of the transition temperature phonons should play an
important role. There also exists strong evidence
that in the insulating phase above $T_c$
charge transport is controlled by the motion of small
magneto--elastic polarons \cite{Myron96}.

%
Notwithstanding the importance of a coupling of lattice, charge
and spin degrees of freedom, the magnetic properties of the
plain DE model are unusual by themselves, which was shown in
finite size calculations
and variational treatments \cite{Zang96,Mueller96,Riera96}. 
It was found that in 1d
the ground state showed unusual odd-even effects,
and that the excitation spectrum even for
ferromagnetic ground states is very unusual \cite{Zang96}.
Similar effects were found in higher dimensions, which could be
understood in terms of the degeneracies of the finite-system fermi
surface\cite{Zang96}. 

In view of the surprising results in \cite{Zang96,Mueller96,Riera96} for
the ground state and low energy excitations of the DE model, it is
necessary to reconsider its finite temperature properties with as
little approximations as possible. We have therefore developed a 
high temperature series expansion (HTSE) for 
the spin-half ferromagnetic Kondo-lattice model
in the large $J_H$ limit for the  internal
energy, the magnetic susceptibility, the compressibility
and the magneto--compressiblity.
The calculations have been done for the simple cubic lattice 
up to order $\beta^{10}$
and, to compare to the simpler mean--field behavior, for the Bethe lattice
(defined as the interior of a Cayley tree with varying
coordination number $q$) up to order   $\beta^{14}$
as a function  of the electron
density $0\le \rho\le 1$. 
The ferromagnetic Kondo-lattice model is defined by the
Hamiltonian

\begin{eqnarray}
{\cal H}&=&-t \sum_{i,j,\sigma}( c^\dagger_{i,\sigma}c_{i,\sigma}^{\phantom{\dagger}} + h.c.)
-J_H\sum_i \vec {S}_i\cdot \vec {\sigma}_{\alpha \beta} c^\dagger_{i,\alpha} c_{i,\beta}^{\phantom{\dagger}} 
\end{eqnarray}

Here $\vec S_i$ is the local spin coupled via the Hund's rule
coupling $J_H$ to the itinerant electron spin 
$\vec {\sigma}_{\alpha \beta} c^\dagger_{i,\alpha} c_{i,\beta}^{\phantom{\dagger}} $.
We are restricting ourselves to local spins of length $s=1/2$.

We are interested in studying this model in the $J_H\to\infty$
limit, in which case there are only five basis states $|s,m>$ 
per lattice site.
Two of these correspond to the up and down spin states of the
local spin when the itinerant electron is absent (a spin--$\frac{1}{2}$
object), and the other
three correspond to the case where the itinerant electron is
present and forms a spin--$1$ object together with the local spin.
These states are 
$|\frac{1}{2},\frac{1}{2}>$, 
$|\frac{1}{2},-\frac{1}{2}>$, 
$|1,1>$,
$|1,0>$ and $|1,-1>$ and can be labelled uniquely
by their $m$--quantum number alone. In order to consider a doped system, 
we add a term $\mu \sum_{i,\sigma} n_{i,\sigma}$ to (1) and adjust
the chemical potential $\mu$
such that these five states are degenerate
(analogous to the HTSE of the $t-J$ model \cite{Singh92}), leaving
the hopping matrix element $t$ as the only energy scale in the
problem. We set $t=1$. 
The action of the effective Hamiltonian in the $J_H \to \infty$ limit 
can be written as 
$
H_{eff}|1,m>_i|\frac{1}{2},m'>_j = -\frac{f_{ij} t}{2}
 [ A_{m,m'} |\frac{1}{2},m+\frac{1}{2}>_i |1,m'-\frac{1}{2}>_j
+B_{m,m'} |\frac{1}{2},m-\frac{1}{2}>_i |1,m'+\frac{1}{2}>_j ]
$,
where $f_{ij}$ represents the fermion sign arising from interchanging
fermions on sites $i$ and $j$ and
$A_{m,m'}=\sqrt{(\frac{3}{2}-m')(1-m)}$,
$B_{m,m'}=\sqrt{(\frac{3}{2}+m')(1+m)}$ \cite{Mueller96}.



HTSEs can be developed for the quantities of
interest by a cluster expansion method \cite{Baker67}. In the
thermodynamic limit, extensive
quantities $X$ are written as a sum over all topologically
distinct graphs $g$, 
$X=N_s\sum_g L(g)\times W(g),$
where $L(g)$ is called the lattice constant of the graph.
 $W(g)$ is called the weight of the graph and is given by
the relation,
$W(g)= X_g-\sum_{g^\prime\subset g} W(g^\prime).$
Here, $X_g$ is
the appropriate quantity calculated for the finite graph and
the sum runs over all subgraphs of the graph $g$.
For a graph with $L$ bonds, the weight $W(g)$ is at least order $\beta^L$.
There is an additional symmetry in our problem, such that
for graphs without closed loops the weight of a graph with $L$
bonds is order $\beta^{2L}$. Thus by including all graphs with up to $L$
bonds one can calculate the desired quantities for the simple cubic
lattice to order $\beta^L$ and for the Bethe-lattice ( which has no
closed loops ) to order $\beta^{2L}$. 
The calculation of the weights is the time consuming step in the 
calculation.
The calculations were
performed on IBM590 workstations and took about 80 CPU days in total.

 After transforming back to the density as the relevant variable
we obtain various thermodynamic quantities as series in $\rho$
(the $e_g$--density), and in $\beta$.
E. g.,  the reduced susceptibility can be written as
\begin{equation}
4\chi / \beta =
1 + \frac{5}{3}\rho + \rho (1-\rho) \sum_m \left( \sum_{l=0}^{m} a_{l,m} \rho^l \right) \beta^{2m}
\end{equation}
The coefficients for this series and for the other quantities
of interest are available on request.
\begin{floating}
\begin{figure}
 \centerline{
 \epsfysize=8.cm \epsfbox{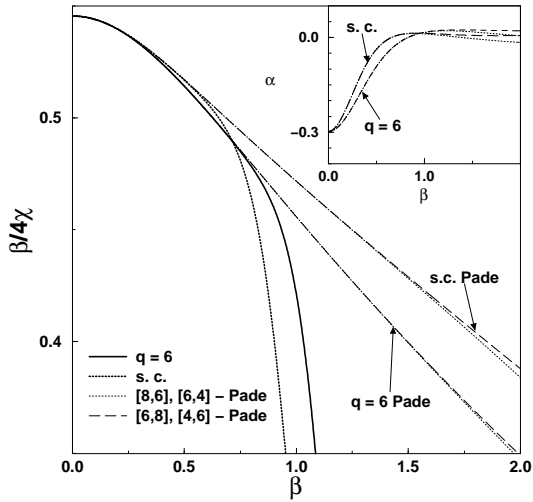}
}
\caption{
Plot of $\beta/ 4\chi$ vs $\beta$ for $\rho=0.5$ on 
the Bethe lattice with $q=6$ and on the s. c. lattice, and the
respective PAs. The inset shows
the second derivative of the respective PAs of $\beta /4 \chi$ 
with respect to $\beta$,
i. e. $\alpha = \partial^2_{\beta} (\beta /4\chi)$. 
}
\label{fig:1}
\end{figure}
\end{floating}

To locate the expected paramagnetic to ferromagnetic transition
temperature $T_c$ let us first consider the susceptibility series.
Although the series for $\chi/\beta$ contains only even powers of
$\beta$, we observe the expected recovery of Curie--Wei{\ss}  behavior
below a typical temperature related to the bandwidth \cite{DeG60}. This
is manifest  in the linearity of $\beta/\chi$ vs. $\beta$
shown in Fig. 1 for both the simple cubic (s. c.) and the 
Bethe lattice with $q=6$ for $\rho=1/2$ (Similar behavior also exists for all
$0<\rho <1$). Unfortunately related to this crossover from
quadratic to linear behavior an unphysical singularity appears
on the imaginary axis. For the series shown in  in Fig. 1 this
singularity is at $\beta_s^{q=6} \approx \pm .94i$, $\beta_s^{s.c.} \approx \pm .85i$ 
obtained from a ratio analysis of the respective series. This
singularity dominates the analytic behavior of the series and
renders conventional methods for extracting $T_c$, like dlog Pad\'es
or Neville tables, unsuccessful. Pad\'e approximants  (PAs) 
actually allow for an extrapolation of the series
beyond its normal radius of convergence (see Fig. 1).
 
For the $q$--coordinated Bethe lattices we can use the expected
mean--field critical behavior
 --- the susceptbility of the ferromagnetic Heisenberg model
for all $T>T_c$ is $\chi = const./(T-T_c)$ ---  to obtain $T_c$ from a fit of
the PAs of $\chi^{-1}$  to a  linear behavior in $T$ (see Fig. 1).
There is, of course, some ambiguity in identifying the region in
temperature where linear behavior holds.
We use the lowest possible range in $T$, where the two
highest order PAs agree 
to obtain an estimate for $T_c$ as a function of $\rho$
and coordination number $q$. 
For $\rho=1/2$, this fit gives e. g. $T_c^{q=6} = .196 \pm .02,
T_c^{q=32} = .485 \pm .023, T_c^{q=256} =1.35 \pm .04$ (The
errorbars are subjective estimates), which indeed follows
the expected scaling with $q$, i. e. $T_c^q \propto q^{1/2}$, reasonably
well, and in retrospect validates the PAs and our
analysis. 
Since for the Bethe lattices we have enough terms, we can actually
also use integral PAs \cite{GeorgeBook}. This yields e. g.
$T_c^{q=6} = 0.21$ for $\rho=1/2$, which agrees within the errorbars with the 
linear fit.
It is worth pointing out that even the highest estimate for $T_c$ from
the  integral PAs, $T_c = 0.22t$, is very low compared to
values obtained by Millis et al.\cite{Millis95}. 
If one takes values of the bandwidth between
$1$--$2.5eV$ one obtains estimates for $T_c$ between $180K$ and $460 K$, which
are well within the experimentally observed range.

The analysis of the series on the simple cubic lattice is more
difficult, since one expects a non--trivial susceptibility
exponent $\gamma >1$, and also because the series is too short to allow
for a sensible analysis using integral PAs. However, just by
comparing the susceptibilities for the $q=6$--Bethe lattice and 
the s.c. (see Fig. 1) one notices that an eventual intercept
with the axis would happen at a larger $\beta$ value for the s.
c. lattice, resulting in an even lower $T_c$.

If one assumes that the critical region is small then one can
analyze the series for the s.c. lattice in a similar way as 
before. However, the resulting estimates for $T_c$ are only 
rough estimates and provide lower bounds (due to $\gamma>1$).
The values at which the PAs still agree can certainly be taken
as upper bounds.
Estimates for $T_c(\rho)$ are shown in Fig. 2.
\begin{floating}
\begin{figure}
 \centerline{
 \epsfysize=8.cm \epsfbox{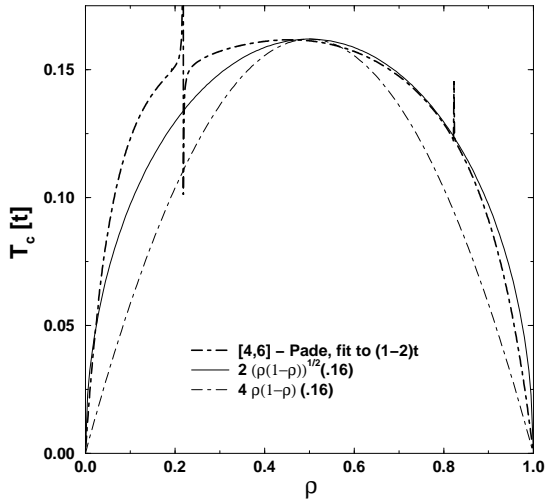}
}
\caption{
Plot of estimates of {$T_c(\rho)$} for the s.c. lattice
obtained from various fit intervals as described in the text. The
spikes arise from
instabilities in the PAs, and have no physical significance. The thin lines
are scaled plots of \protect{$\rho(1-\rho)$} (dash--dot) and 
$(\rho(1-\rho))^{1/2}$ (solid).}
\label{fig:3}
\end{figure}
\end{floating}
The values are consistently about $25\%$ 
lower than the $q=6$ Bethe lattice estimates. 

The small curvature in the inset of Fig. 1 
may be indicative of nontrivial corrections.
It is not unnatural to assume that at least at
$\frac{1}{2}$--filling the system consists of ferromagnetically 
coupled spin--$\frac{1}{2}$s and spin--$1$s, and if one 
assumes some kind of charge ordering (e.g. in an AB--structure)
this would give rise to  behavior reminiscent of 
a ferromagnetic ``{\it ferrimagnet}''. For such a 
system 
\begin{equation}
4\chi = [(1+\frac{5}{3}\rho)T+2\sqrt{\frac{8}{3}\rho(1-\rho)}T_c]/(T^2-T_c^2)
\label{ferri}
\end{equation}
within mean-field theory \cite{whitebook}.
The density dependence of (\ref{ferri}) is obtained
in a mean--field picture. We imagine that large spins
of size $2S_1=(1-\rho)$ (the empty sites) and size
$2S_2=8\rho/3$ (the occupied sites) are coupled
via a ferromagnetic $J_{12}$. Possible other couplings,
like $J_{11}$ and $J_{22}$ are zero in our case of the
DE limit, i. e. there are no induced spin--spin interactions
between two empty or between two occupied sites.
For $\rho=1/2$ the slope of $\chi$  actually
agrees with (\ref{ferri}) (for the Bethe lattices). 
However, the actual difference from the 
value for a Heisenberg model are small, and for
other values of $\rho$ the agreement of our series with (\ref{ferri})
becomes worse, although still showing the trend contained in (\ref{ferri}).
Ferrimagnetic mean--field theory also predicts the concentration
dependence
of $T_c = J_{12} \sqrt{6\rho(1-\rho)}/9$ as opposed to the expected $T_c(\rho) \propto 
\rho(1-\rho)$, see e. g. \cite{Millis95,Varma96}. Apart from the unphysical spike
around $\rho=0.3$ the agreement of estimation from our HTSE
with the prediction from ferrimagnetic mean--field theory is remarkable.

This still leaves open the question of emerging structure in
the charge degrees of freedom. One way of looking at this within
our framework is to consider the compressibility
defined as $\kappa = ({\partial\rho\over\partial\mu})/\rho^2$ and shown in Fig. 3.
\begin{floating}
\begin{figure}
 \centerline{
 \epsfysize=8.cm \epsfbox{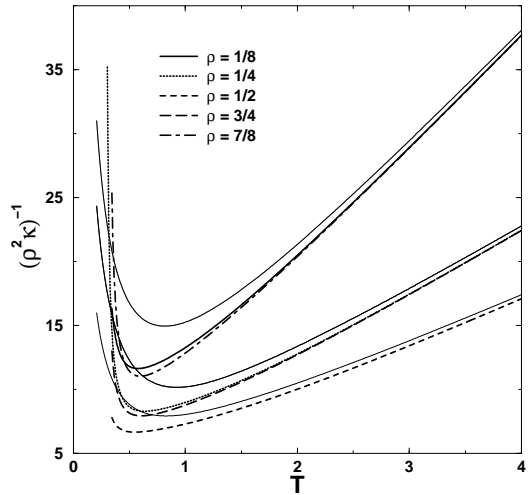}
}
\caption{
The inverse compressibility $(\rho^2 \kappa)^{-1}$ as a function of
temperature $T$. Shown are the [4,6]--PAs to the 10$^{th}$ order
series on the s. c. lattice. The thin lines through each pair depict the
values of $(\rho^2\kappa)^{-1}$ from the HTSE for free 
spinless fermions on a s.c. lattice
at the corresponding densities $\rho$.
}
\label{fig:4}
\end{figure}
\end{floating}
Again we find a quasi--linear behavior below a certain temperature.
By comparing with a HTSE for spinlesss fermions on the s. c. lattice
we see that differences in the compressibility between the DE model
and noninteracting spinless fermions
are small and only become relevant at low temperatures.
This means that the DE model at high temperatures behaves pretty
much like a system of non--interacting spinless fermions 
at least with respect to charge excitations.
We do not see any evidence for unusual behavior in $\kappa$ that
would be indicative of some kind of charge ordering.
However, this is not conclusive since  one would expect unusual behavior
to occur at a specific wavevector related to the filling,
and hence it is not surprising to see nothing in our
$q=0$ compressibility. 

To connect more closely to the CMR effect it would be necessary
to calculate the magnetoresistance (MR) 
$\rho_c(h)/\rho_c(h=0)-1$, where $\rho_c$ is the resistivity and $h$ the applied 
magnetic field. The MR arises from two sources. One is a 
change in the scattering time due to the magnetic field, which
modifies the dynamical properties, the other is the dependence of
the bandwidth on the magnetic field, which is a static effect.
The latter is large in the DE model as the
effective bandwidth is strongly dependent on the alignment of
the local spins. In a fermi-gas, the compressibility is proportional
to the effective mass. Thus, we can capture  this bandwidth effect in
the magneto--compressibility (MC)  $\kappa(h)/\kappa(h=0)-1$. 
In Fig. 4, we show the MC for 
various densities as a function of inverse temperature. As expected,
and desired, it is always negative. For intermediate $\rho$ 
we observe saturation of the MC  around $\beta=1$ corresponding to 
the appearance of linear behavior of $\chi^{-1}$. This does not 
agree with the behavior observed in the MR of the real materials.
Also the concentration dependence of the MC is inconsistent
with that of the MR; the higher the concentration of $e_g$ electrons the
larger is the MC. This can be understood  by noting that for larger
$\rho$ there is effectively {\it more} spin to be polarized by the
magnetic field. One way to resolve this discrepancy would be to 
assume that the dynamic effects in the MR are more important than the
static ones captured within the MC.

\begin{floating}
\begin{figure}
 \centerline{
 \epsfysize=8.cm \epsfbox{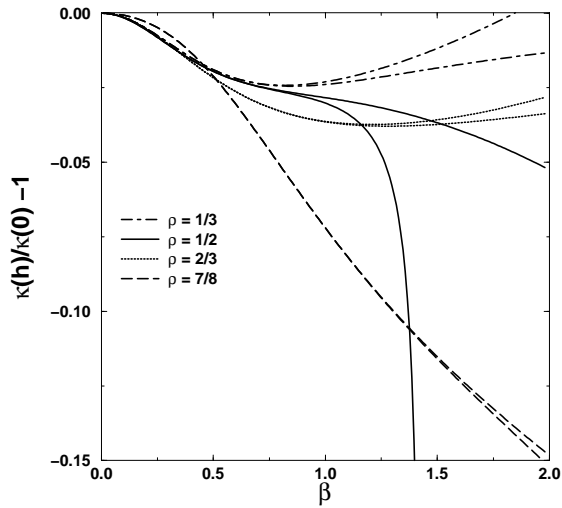}
}
\caption{The MC  as defined in the 
text as a function of
$\beta$. Shown are
two PAs (upper:[4,6] lower: [6,4]) for
the respective densities $\rho$.
}
\label{fig:5}
\end{figure}
\end{floating}
In this letter we have provided the first exact results for the
ferromagnetic Kondo lattice model in terms of a HTSE. Although
we have used a smaller spin value than is realized in the
CMR materials, the $t_{2g}$ spin length is set to $1/2$ instead of $3/2$,
this should not affect $T_c$ much as, to leading order for
$J_H\to\infty$, the spin-length drops out of the problem.
The analytic structure of the resulting series is governed 
by a singularity on the imaginary axis, which we indentify
as being responsible for the crossover from the high--T range to
the physically interesting temperature range below the
bandwidth. This behavior is probably also typical of other 
itinerant systems, although the DE model in the $J_H \to \infty$
limit is unusual, because the local moment $\sum_i <(S_i^z)^2>$ is a 
constant, and does not depend on temperature.
Our estimates for the magnetic transition temperature are close
to the experimentally observed ones. Mean--field theory, as
mimicked by the Bethe lattice series, overestimates $T_c$ by about
$25\%$.
There are also some unusual
features  associated with the charge degrees of freedom as indicated by 
the slightly ferrimagnetic behavior, but our calculations are not
sufficient to study long-range charge order. Nevertheless, the
tendencies towards charge ordering should imply 
a large sensitivity of the CMR materials towards Jahn--Teller
ordering effects. 
In further work we will investigate the
wave-vector dependence of the compressibility to address this question.
 Our results for the magneto--compressibility indicate that
dynamic effects are responsible for the CMR effect, and that a
static description may be inadequate.
\begin{flushleft}
{\bf Acknowledgements}
\end{flushleft}
We thank A. R. Bishop, S. Trugman, G. Baker, Jr., and D. Wallace
for many illuminating discussions. This work is supported in part
by NSF grant number DMR-9616574.

Thanks a lot. 

Best Regards, Rajiv
and by the University of California
campus laboratory collaboration (CLC).

\end{document}